\documentclass[12pt]{jstyle}
\usepackage{amsmath,amssymb,epsfig,bm,amsfonts,yfonts}

\newcommand{\be}{\begin{equation}}
\newcommand{\ee}{\end{equation}}
\newcommand{\beqa}{\begin{eqnarray}}
\newcommand{\eeqa}{\end{eqnarray}}
 \newcommand{\bn}{\begin{enumerate}}
\newcommand{\en}{\end{enumerate}}

\def\bra#1{\left\langle #1\right|}
\def\eeq{\end{equation}}

\def\ket#1{\left| #1\right\rangle}

\def\Tr{\mathop{\rm Tr}}

\relax

\title{Stirring Strongly Coupled Plasma}

\author{Kazem Bitaghsir Fadafan${}^{\,1}$, Hong Liu${}^{\,2}$, Krishna Rajagopal${}^{\,2}$ and
Urs Achim Wiedemann${}^{\,3}$\\
\vspace{0.1in}

${}^{\,1}$Physics Department, Shahrood University of
Technology, Shahrood, Iran
\vspace{0.1in}

${}^{\,2}$Center for Theoretical Physics, MIT,
Cambridge, MA 02139, USA
\vspace{0.1in}

${}^{\,3}$Department of Physics, CERN, Theory Division, CH-1211 Geneva 23
\vspace{0.1in}

E-mail addresses: {\tt bitaghsir@shahroodut.ac.ir, hong\_liu@mit.edu, krishna@ctp.mit.edu, Urs.Wiedemann@cern.ch}
}

\abstract{
We determine the energy it takes to move a test quark along a circle of
radius $L$ with angular frequency $\omega$ through
the strongly coupled plasma of ${\cal N}=4$ supersymmetric Yang-Mills (SYM) theory.
We find that for most values of $L$ and $\omega$
the energy deposited by stirring the plasma in this
way is governed either by the drag force acting on
a test quark moving through the plasma in a straight
line with speed $v=L\omega$ or by the energy radiated
by a quark in circular motion in the absence of any plasma, whichever
is larger.  There is a continuous crossover from the drag-dominated
regime ($\omega\lesssim\pi T(1-v^2)^{3/4}$, meaning $\omega\lesssim\pi T$
and $L$ small enough)
to the radiation-dominated regime ($ \omega\gtrsim\pi T(1-v^2)^{3/4}$).
In the crossover regime we find evidence for significant destructive interference
between energy loss due to drag and that due to radiation as if in vacuum.
The rotating quark thus serves as a model system in which the relative
strength of, and interplay between, two different mechanisms of parton
energy loss is accessible via a controlled classical gravity calculation.
We  close by speculating
on the implications of our results
for a quark that is moving through the plasma
in a straight line while decelerating, although in this
case the classical calculation breaks down at the same
value of the deceleration at which the radiation-dominated regime
sets in.
}

\preprint{MIT-CTP-3979, CERN-PH-TH/2008-194}

\begin{document}
\def\vev#1{\langle#1\rangle}
\def\ov{\over}
\def\le{\left}
\def\ri{\right}
\def\ha{{1\over 2}}
\def\lam{{\lambda}}
\def\Lam{{\Lambda}}
\def\al{{\alpha}}
\def\ket#1{|#1\rangle}
\def\bra#1{\langle#1|}
\def\vev#1{\langle#1\rangle}
\def\det{{\rm det}}
\def\tr{{\rm tr}}
\def\Tr{{\rm Tr}}
\def\NN{{\cal N}}
\def\th{{\theta}}

\def\Om{{\Omega}}
\def \th{{\theta}}

\def \lam {\lambda}
\def \om {\omega}
\def \ra {\rightarrow}
\def \ga {\gamma}
\def\sig{{\sigma}}
\def\ep{{\epsilon}}
\def\apr{{\alpha'}}
\newcommand{\p}{\partial}
\def\LL{{\cal L}}
\def\HH{{\cal H}}
\def\GG{{\cal G}}
\def\TT{{\cal T}}
\def\CC{{\cal C}}
\def\OO{{\cal O}}
\def\PP{{\cal P}}
\def\tir{{\tilde r}}

\newcommand{\bea}{\begin{eqnarray}}
\newcommand{\eea}{\end{eqnarray}}
\newcommand{\nn}{\nonumber\\}

%%%%%%%%%%%%%%%%%%%%%%%%%%%%%%%%%%%%%%%%%%%
\section{Introduction}
Parton energy loss in hot plasmas
has been studied intensely in recent years~\cite{jetquenchrev}.
Experimentally, the phenomenon can be accessed
by measuring the remnants of jets
that are quenched by the hot and dense matter produced
in relativistic heavy ion collisions at RHIC~\cite{RHIC} or at the LHC~\cite{LHC}.
The modification of quenched jets provide one of the best tools
for constraining properties of the matter produced
in heavy ion collisions.
Furthermore, the energy lost by the quenched jet
provides a localized perturbation of the surrounding
matter that is expected to give rise to characteristic
collective phenomena
such as cone-like momentum flow patterns~\cite{CasalderreySolana:2004qm}.
These are searched for in the data.

In the high projectile-energy limit, the energy loss
of a relativistic parton is dominated by gluon bremsstrahlung
and calculations based upon perturbative QCD are reliable~\cite{jetquenchrev}.
These calculations describe the destructive interference between
the vacuum radiation of a virtual parton and the
additional QCD bremsstrahlung radiation
that is sensitive to the acceleration of the colored projectile
moving through the medium~\cite{jetquenchrev}.
At lower projectile energies
other mechanisms including collisional energy loss via
elastic interactions may contribute to parton energy loss.
And, even in the high projectile-energy limit the interactions of the partonic
jet fragments with the medium involve small momentum transfers, of order
the temperature if the medium is in equilibrium, and are thus expected
to be governed by a nonperturbatively strong coupling constant.
However,
strong coupling calculations of dynamic processes
in QCD are notoriously difficult. This has motivated
parton energy loss calculations in a class of supersymmetric theories,
for which the AdS/CFT correspondence~\cite{AdS/CFT} provides
powerful techniques for doing calculations at strong coupling.
Two different strategies have been pursued:

In one approach, one identifies the nonperturbative properties
of the strongly coupled medium that must be
put  into the otherwise perturbative
QCD energy loss calculations in terms of a light-like Wilson loop~\cite{Wiedemann:2000za},
which is then calculated
with the help of the AdS/CFT correspondence~\cite{Liu:2006ug,Liu:2006he}.
The range of validity of this calculation is
limited to very energetic partonic projectiles, where the dominant energy loss mechanism
is indeed radiative and the perturbative calculation (with a nonperturbative input
that plays the same logical role that parton distribution functions do
in deep inelastic scattering) is valid.

In the other approach, one assumes that the initial production of hard
partons is perturbative\footnote{This must be the case since if the initial hard
scattering were also described within the strongly coupled theory, one would
observe hedgehog-like 
hard-scattering events rather than the jet-like events
seen at RHIC~\cite{Hatta:2007cs,Hofman:2008ar}.}
but then treats the parton energy loss process as if interactions
at all relevant scales are nonperturbatively strong and therefore
formulates the entire energy loss
calculation in the gravity dual of a strongly coupled supersymmetric gauge theory.
The simplest problem to consider is the force required to move a projectile
through the plasma at some constant velocity~\cite{Herzog:2006gh,Gubser:2006bz}.
If the projectile velocity is sufficiently small, the gravity description is
in terms of a classical string trailing ``down'' into
the five-dimensional spacetime behind the quark moving along the
four-dimensional boundary.  This classical calculation breaks down
at high enough velocities because the force required to maintain
the quark at constant velocity becomes large enough that quark
antiquark pair production is unsuppressed~\cite{CasalderreySolana:2007qw}.  In the velocity regime
in which this calculation is reliable, it yields the result that the
force required to keep the quark moving is proportional to
the quark momentum, meaning that energy loss occurs
via drag~\cite{Herzog:2006gh,Gubser:2006bz}.
Quantum fluctuations of the string trailing behind the quark translate
into fluctuations in the momentum of the
quark~\cite{CasalderreySolana:2006rq,Gubser:2006nz,CasalderreySolana:2007qw}.
It is an open
question, however, how to make a quantitative
translation from these results to the physically relevant setting
in which there is no external force acting on the energetic
colored projectile, which is therefore decelerating.
This deceleration opens the possibility of additional
mechanisms of parton energy loss, since at least in vacuum
we know that it would  lead to radiation.  And, it raises the possibility
of interference between medium-induced energy loss (in this case drag)
and deceleration-induced energy loss.
Motivated by these questions, which remain open, we have
found a different simple problem to consider in which, within
the regime of validity of the classical trailing string calculation,
we find that energy loss can either be dominated  by drag or can
be as if the quark were radiating in vacuum, and behaves
as if energy loss via these two mechanisms interfere
destructively.

In the present work, we determine the energy needed to move a heavy test quark along
a circle of radius $L$ with angular frequency $\omega$ through the strongly coupled
plasma of ${\cal N}=4$ supersymmetric Yang-Mills (SYM) theory.
We are interested in this problem since it provides a novel
perspective on several of the open questions mentioned above. Because
in vacuum a rotating colored particle would emit synchrotron radiation,
studying a particle moving in a circle with constant angular  velocity
through a strongly coupled plasma is well suited to studying the
relative strength of, and interplay between, radiation and medium-induced
energy loss, as a function of $\omega$ and $L$.
It will also allow us to compare radiation in medium and
in vacuum. Also, in contrast to linear motion with acceleration, the case of
rotation at constant angular velocity can
be formulated as an essentially time-independent problem
even though it includes acceleration.
Finally, stirring a strongly coupled plasma
by a rotating quark is a well-localized source of plasma perturbations
whose propagation and
dissipation is described in the dual gravity theory by the trailing string.
We leave the translation from the trailing string description to the
corresponding stress-energy
tensor describing the disturbance of the gauge theory plasma to future work.

Our paper is organized as follows: In Section~\ref{sec2}, we calculate the shape of the
trailing string spiralling below the quark at its endpoint, which is being
pulled with constant
angular velocity along a circle within strongly
coupled ${\cal N}=4$ SYM plasma. This is the gravity dual of
a rotating heavy quark and the disturbance that the
rotating quark creates by depositing energy in the plasma.
In Section~\ref{sec3}, we determine the corresponding energy
loss and compare the result to expectations
if the energy loss were due to acceleration-induced radiation in
the absence of any medium and to expectations if the energy
loss were due to drag in the absence of any acceleration.
We find that in regimes in which one of these is much larger
than the other, the larger one is a good approximation to our
result.  Where both expectations are comparable, our result
is less than their sum, indicating destructive interference.
By comparing the shape of the trailing string in medium to
that in vacuum, we show that even
in the regime in which the energy loss
behaves precisely
as if the rotating quark were emitting synchrotron radiation
in vacuum, the disturbance of the plasma
due to the deposited energy is not the same as the radiation pattern
for synchrotron radiation in vacuum.
In Section~\ref{sec4} we return to the case of linear motion, which
motivated our investigation.  We show that the classical
calculation breaks down just at the velocity at which
energy loss due to deceleration-induced
radiation becomes comparable to that due to drag.
This is in contrast to the case of circular motion, where we
find a wide range of parameters $\omega$ and $L$ where
the classical calculation
is reliable and acceleration-induced radiation dominates over drag.
We nevertheless close by using our results to speculate about the
role of radiation for the case of linear motion.

%%%%%%%%%%%%%%%%%%%%%%%%%%%%%%%%%%%%%%%%%%%%%%%
\section{A Rotating Quark and a Spiralling String}
\label{sec2}

 \subsection{Formulation of the problem}

 We shall consider a heavy test quark moving through the
 strongly coupled plasma of ${\cal N} = 4$ SYM with temperature $T$.
 We assume that there is some external agent exerting a suitable force
 on the quark such that it moves along a circle of radius $L$
 at a  constant angular frequency $\omega$.
 The quark has a constant speed
\begin{equation}
    v = L\omega\ ,
    \end{equation}
and it has a constant acceleration perpendicular to its direction of motion,
\begin{equation}
	a = \omega v = \omega^2\,L\, .
\end{equation}
We wish to compute $dE/dt$, the energy lost per unit time by the quark as it
stirs the plasma.  $dE/dt$ is the energy expended by the external agent
moving the quark, and it is also the energy dumped into the plasma
that is being stirred.  We shall do the computation using the gravity
dual of ${\cal N}=4$ SYM, following  the basic logic first developed
in the study of a test quark being moved along a
straight line at constant speed through the plasma~\cite{Herzog:2006gh,Gubser:2006bz}.
Although we do not know of a physical realization of the problem
that we solve, it is instructive because, as we
shall see, it exhibits a crossover between
a regime in which $dE/dt$ is dominated by the drag force experienced
by a quark moving in a straight line with speed $v$ to a regime in which
$dE/dt$ is dominated by the radiation that arises by virtue of the
acceleration $a$.

${\cal N}=4$ SYM theory
is a supersymmetric gauge theory characterized by two parameters: the
rank of the gauge group $N_c$ and the 't Hooft coupling $\lambda = g^2_{\rm YM} N_c$,
where $g_{\rm YM}$ is the gauge coupling.   The theory is conformal, meaning
that $\lambda$ is a parameter that we can choose.  If we choose $\lambda$ large,
at $T\neq 0$ this theory describes a strongly coupled plasma.
According
to the AdS/CFT correspondence~\cite{AdS/CFT},
this gauge theory is equivalent to
Type IIB string theory in AdS$_5 \times S_5$ spacetime,
with the curvature radius
$R$ of the Anti deSitter (AdS) space
and the string tension $1/(2\pi\alpha')$ related to the 't Hooft coupling by
\begin{equation}
\sqrt{\lambda}=\frac{R^2}{\alpha'}\ .
\label{LambdaAlpha}
\end{equation}
The gauge theory can be thought of as living on the boundary of AdS$_5$.
If we take $N_c\rightarrow\infty$
at fixed $\lambda$ and then take $\lambda$ large,
${\cal N}=4$ SYM theory has
a gravity dual: it is described by classical supergravity on  AdS$_5 \times S_5$.
Nonzero temperature $T$ in the gauge theory corresponds to replacing
the AdS$_5$ spacetime in the gravity dual by a 5-dimensional AdS black hole,
with the metric
\begin{eqnarray}
	ds^2 &=& -f(\tilde u)\,
	dt^2 + \frac{\tilde u^2}{R^2} \left(d\tilde\rho^2 + \tilde\rho^2 d\varphi^2 + dx_3^2 \right)
		+ \frac{d\tilde u^2}{f(\tilde u)} \, ,
		\label{eq1}\\
		f(\tilde u) &\equiv& \frac{\tilde u^2}{R^2} \left(1 - \frac{\tilde u_h^4}{\tilde u^4} \right)\, .
		\label{eq2}	
\end{eqnarray}
Here, the bulk coordinate in the 5th dimension is $\tilde u$ and $\tilde u_h$ is
the Schwarzschild radius at which we find the horizon of
the black hole.  The temperature in the gauge theory is equal to the Hawking temperature
of the AdS black hole in the gravity dual, namely
\begin{equation}
	T = \frac{\tilde u_h}{\pi R^2}\, .
\end{equation}
In (\ref{eq1}), we have written two of the three spatial dimensions using radial coordinates
$(\tilde\rho,\varphi)$, with the third spatial coordinate being $x_3$.
The coordinates $\tilde u$ and $\tilde \rho$ have dimensions of length; we
shall soon replace them by dimensionless coordinates, and drop the tildes.

In the gravity dual, a heavy test quark corresponds to an open string
with one end-point on a
D3-brane at $\tilde u = \infty$. The string hangs down
into the bulk, extending towards
the black hole horizon at $\tilde u=\tilde u_h$.
In order to use
the gravity dual description of ${\cal N}=4$ SYM to determine $dE/dt$ for a quark moving
in a circle, we must first (in this Section) find the worldsheet of the string spiralling downward from
the quark moving on a circle at $u=\infty$ and then (in
the next Section) calculate the energy flowing
down the string.
We choose to parameterize the two dimensional worldsheet
of the rotating string
$X^\mu(\tau,\sigma)$ according to
\begin{equation}
	X^{\mu}(\tau, \sigma) = \left(t=\tau,\tilde\rho=\tilde\rho(\tilde u),
	\varphi=\omega\tau+\theta(\tilde u),x_3=0,\tilde u=\sigma \right)\,.
	\label{eq3}
\end{equation}
Here and throughout, indices $\mu$ and $\nu$ run over five dimensions,
$\mu, \nu \in  (t, \tilde\rho, \varphi, x_3, \tilde u)$. Below,  indices $a$, $b$ run
over the two dimensions of the
worldsheet, $a,b \in (\sigma\, ,\tau)$.
In order to describe the rotating quark that we wish to analyze, the
end-point of the string on the D3-brane must satisfy the boundary conditions
\begin{eqnarray}
	\tilde{\rho}(\infty) &=& L\, ,
	\nonumber \\
	\theta(\infty) &=& 0\, .
\end{eqnarray}
In writing the parameterization (\ref{eq3}), we have made use of the fact that the quark
is in circular motion at a constant angular velocity, and has been
moving in this way for all time. The entire
time-dependence of the string therefore consists
of a global rotation with the azimuthal angle $\phi$
increasing with time according to $\phi(\tau) = \omega \tau$
plus a $\tau$-independent, but $\tilde u$-dependent, function.
This function $\theta(\tilde u)$,
together with the function $\tilde\rho(\tilde u)$ that specifies how the radius
of the string depends on $\tilde u$,  provide a complete
specification of  the shape of the spiralling string
dangling down from the rotating quark.  In order to visualize this
set-up, it may be helpful to look at the example solution in Fig.~\ref{fig2}.

We find explicit solutions $\theta(\tilde u)$ and $\tilde\rho(\tilde u)$ for
the shape of the spiralling string
by minimizing the Nambu-Goto action
\begin{eqnarray}
	S &=& - \frac{1}{2\pi\alpha'}\int d\tau\, d\sigma\,  \sqrt{-{\rm det} g_{ab} } \equiv - \frac{1}{2\pi\alpha'}\int d\tau\, d\sigma \, {\cal L}  \, ,
		\label{eq4}
\end{eqnarray}
where
\begin{equation}
g_{ab}\equiv G_{\mu\nu} \partial_a X^\mu \partial_b X^\nu
\end{equation}
is the induced metric on the worldsheet and $G_{\mu\nu}$
is the spacetime metric (\ref{eq1}).
We shall find it convenient to describe the shape
of the string with the dimensionless  variables
\begin{eqnarray}
	 u &\equiv& \frac{\tilde u}{\tilde u_h} \ ,
		\label{eq12} \\
	\rho &\equiv& \frac{\tilde u_h}{R^2} \tilde\rho\, ,
	           \label{eq13}
\end{eqnarray}
and specify the motion of the quark with
the dimensionless parameters
\begin{eqnarray}
\ell&\equiv&\frac{\tilde u_h}{R^2}\, L = L\pi T \ ,\label{eq14a}\\
\textswab{w}&\equiv&\frac{R^2}{\tilde u_h}\, \omega
= \frac{\omega}{\pi T}\ .
		\label{eq14}
\end{eqnarray}
Note that
\begin{equation}
v=L\omega=\ell \textswab{w}
\end{equation}
and note that the dimensionless
acceleration of the quark at $u=\infty$ is given by
\begin{equation}
\textswab{a} = \textswab{w} v = \textswab{w}^2\ell
= \frac{a}{\pi T}\ .
\end{equation}
The Lagrangian
now takes the form
\begin{eqnarray}
	{\cal L} &=&
	\sqrt{ \left(u^4 - \rho^2 \textswab{w}^2 u^4 - 1\right)
		\left(\rho'^2 + \frac{1}{u^4-1}\right) +
	         \rho^2 \left(u^4-1\right)\, \theta'^2}  \, .
	         \label{eq15}
\end{eqnarray}
Here, the prime denotes a derivative with respect to the dimensionless bulk variable $u$, e.g.
$\rho' \equiv \partial \rho/\partial u$.
We then obtain the equations of motion
\begin{eqnarray}
	\partial_u \frac{\partial {\cal L}}{\partial \rho'} -  \frac{\partial {\cal L}}{\partial \rho}&=& 0\, ,
	\label{eq16}\\
	\partial_u\frac{\partial {\cal L}}{\partial \theta'} &=& 0
	\label{eq17}
\end{eqnarray}
from the Lagrangian (\ref{eq15}).
These are the equations that determine
$\rho(u)$ and $\theta(u)$ and hence
the shape of the spiralling string trailing below the rotating quark.

To solve the equations of motion (\ref{eq16}), (\ref{eq17}), we first focus on the angular
dependence. The Lagrangian depends on $\theta'$ but
not on $\theta$, so there is a constant of the motion that we shall define as
\begin{equation}
	\Pi  \equiv - \frac{\partial {\cal L}}{\partial \theta'}
		= \frac{\theta'\, \rho^2\, (u^4-1)}{\cal L}\, .
		\label{eq18}
\end{equation}
$\Pi$ will play a central role in our analysis.
We shall see that for a given $\textswab{w}$ the spiralling string solutions
can be specified either by giving the radius $\ell$ of the motion of the
quark at $u=\infty$ or by giving $\Pi$.  And, we shall see in Section 3
that the energy lost to the medium by the rotating quark is proportional
to $\Pi$.
Starting from (\ref{eq18}), we can solve for $\theta'(u)$ as a function of $\rho(u)$, obtaining
\begin{eqnarray}
	{\theta'}^2 = \frac{\Pi^2\, \left(u^4-\rho^2\textswab{w}^2
u^4-1\right) \left(\rho'^2
	+ \frac{1}{u^4-1}\right)}{\rho^2\, \left(u^4-1\right)\, \left[\rho^2\, \left(u^4-1\right)- \Pi^2\right]}\, .
	\label{eq19}
\end{eqnarray}
We now see that we can use (\ref{eq19}) to eliminate $\theta'^2$ from
(\ref{eq16}),  obtaining
an equation of motion for $\rho(u)$
in which the $\theta$-dependence manifests
itself only through the presence of the constant $\Pi$.
This equation can be written in the form
\begin{equation}
\rho'' + \frac{2 u^3 \rho \rho' -1}{\rho^2(u^4-1)-\Pi^2}\,\rho \left(\rho'^2(u^4-1) + 1\right)
+ \frac{2u^3 \rho\rho'\left(1-\rho^2\textswab{w}^2+\rho^2\rho'^2\textswab{w}^2\right)
+\rho'^2(u^4-1)+1}{(1-\rho^2\textswab{w}^2)\rho u^4 - \rho} = 0\ .
\label{rhoEOM}
\end{equation}
After first solving the
differential equation (\ref{rhoEOM}) to obtain $\rho(u)$, we will then be able to
integrate (\ref{eq19}) to
obtain $\theta(u)$.

We note that for $\Pi = 0$, the angular equation of motion is trivial, $\theta'(u)=0$, and
the radial equation of motion (\ref{rhoEOM}) allows for another set of physical solutions with
any radius $\ell < 1/\textswab{w}$. These solutions have been studied as
a model of rotating mesons~\cite{Peeters:2006iu,Burikham:2007kp,Antipin:2007mz}. Their radial
shape $\rho(u)$ decreases with decreasing
$u$, reaches a turning point where $\rho(u_{\rm turn}) = 0$,  $\rho'(u_{\rm turn}) = 0$,
and then rises back to $u=\infty$, which is reached on the same circle with radius $\ell$
at $\theta_{\rm end} = \theta_{\rm start} + \pi$. These mesonic string configurations do
not experience any energy loss within the supergravity
approximation. In the present work, we focus solely on spiralling string
solutions with $\Pi \neq 0$, noting from (\ref{eq19}) that no solutions that pass through
a turning point with $\rho=0$ can exist if $\Pi\neq 0$.

%%%%%%%%%%%%%%%%%%%%%%%%%%%%%%%%%%%%%%%%%%%%%
\subsection{One special point on the spiralling string}
Before implementing the procedure that
we have just described for finding the string worldsheet, it is helpful to first
consider some general properties of the solutions that  we are looking for.
The rotating quark corresponds to a string which starts
at the boundary $u=\infty$
at a radius $\rho(\infty) = \ell$,
and which hangs down into the bulk and approaches the black hole horizon.
At the boundary, the numerator in (\ref{eq19}) is positive
because $v=\ell\textswab{w}<1$.
At each position $u$ in the bulk, the string will be subject
to a centrifugal force.
This implies that $\rho(u) > \ell$ for all values of $u<\infty$ in the bulk.
Consequently, once the string reaches the horizon, $u=1$,
the factor $(u^4-\rho^2\textswab{w}^2 u^4 -1)$
in the numerator in (\ref{eq19}) must have changed sign and become negative.
The numerator changes sign at the point $(u_c,\rho_c)$ where the
string solution crosses the curve
\begin{eqnarray}
	\rho_{\rm light}(u,\textswab{w})
= \frac{1}{\textswab{w}}\, \sqrt{1-\frac{1}{u^4}}\, ,
	\label{eq42}
\end{eqnarray}
which is the curve at which the local velocity of the string
%$v(u)=
$\textswab{w}\,\rho (u)$
becomes equal to the local speed of light at this depth in the bulk as viewed
from infinity, namely
$c_{\rm light}(u) =  \sqrt{ 1 - \frac{1}{u^4}}$.
However, $\theta'^2$ itself must not change sign, since $\theta'$ must be real.
This can only be the case if the denominator of (\ref{eq19}) changes sign
at the same point that the numerator does, which implies
that $\rho_c^2(u_c^4-1)=\Pi^2$. This condition uniquely
specifies the point at which the
solution describing a string worldsheet with a particular value of the
constant $\Pi$ must cross the curve $\rho_{\rm light}(u,\textswab{w})$,
yielding
\begin{eqnarray}
  u_c &=& \sqrt{ \frac{\Pi\, \textswab{w}}{2}
+ \frac{1}{2} \sqrt{4+\Pi^2\, \textswab{w}^2} }\, ,
	\label{eq43}\\
	\textswab{w} \rho_c &=&
	\sqrt{\frac{2\Pi \textswab{w}}{\Pi\, \textswab{w}
+  \sqrt{4+\Pi^2\, \textswab{w}^2}}}\, .
	\label{eq44}
\end{eqnarray}
The rotating string solution starts at $u=\infty$, $\rho=\ell$,
passes through $(u_c,\rho_c)$
and, we shall see,
approaches the black hole horizon located at $u=1$ at a finite value
of $\rho$ that
is greater than $\rho_c$ which in turn is greater than $\ell$.
The point $(u_c,\rho_c)$ is completely analogous to one that occurs on the
worldsheet of the string that trails behind and below a quark
moving in a straight line at constant speed~\cite{Herzog:2006gh,Gubser:2006bz}.
As in that context, it demarcates the location of a horizon
in the worldsheet metric $g_{ab}$.  It
separates the upper part of the string with $u>u_c$
which moves slower than the
local velocity of light  from the lower part of the string with $u< u_c$
whose local velocity exceeds that of light.
By inspection
of (\ref{eq15}) and (\ref{eq42}) we notice that $\rho''$ appears in the
equation of motion (\ref{eq16}) multiplied by a factor which vanishes
at $(u_c,\rho_c)$.  This means that the equation of motion
itself determines $\rho'(u_c)$, independent of any feature of the
solution at any point away from $(u_c,\rho_c)$.\footnote{It is possible
to show that $\rho'(u_c)$ is given by the negative root of the equation 
 \begin{equation}
 v_c \textswab{w}^2 + \frac{4 \sqrt{1 - v_c^2} + \textswab{w}^2 }{ (1 - v_c^2)^\frac{1}{4}} \rho'(u_c) 
 - \frac{ v_c^3 }{ 1-  v_c^2} \rho'(u_c)^2 =0
  \end{equation}
  where $v_c\equiv\rho_c \textswab{w}$.
}
Therefore, small fluctuations of the string at $u>u_c$ are causally disconnected
from those at $u<u_c$,  meaning in particular
that the lower part of the string is disconnected from its
endpoint on the D3-brane
at the boundary~\cite{Gubser:2006nz,CasalderreySolana:2007qw}.
It has been argued recently that
the lower part of such a string
represents the ``gluon cloud'' that the
quark has lost to the medium it is moving through (i.e. the
energy it has ``radiated'', using this phrase to encompass
energy loss due to drag) while the
upper part of the worldsheet represents the color fields carried
within the quark wave
function~\cite{Dominguez:2008vd}.

%%%%%%%%%%%%%%%%%%%%%%%%%%%%%%%%%%%%%%%%%%%%%%%
%
%%%%%%%%%%%%%%%%%%%%%%%%%%%%%%%%%%%%%%%%%%%%%%
\subsection{Radial dependence of the spiralling string}

\FIGURE[t]{
\includegraphics[width=18cm,angle=0]{rhoplot.eps}
\caption{The radial dependence $\rho(u)$ of string worldsheets
describing the spiralling strings hanging down from rotating quarks
with three choices of angular velocity, $\textswab{w} = 0.05$,
$\textswab{w} = 0.5$ and $\textswab{w}=5.0$,
and a set of values for the
constant of the motion $\Pi$. The figure shows that small (large) values of
$\Pi$ correspond to small (large) values of $\rho(\infty)=\ell$, the
radius of the circle along which the quark at the boundary is moving.
Each solution reaches the horizon $u=1$ at a finite $\rho(1)$.
We also show
the curve $\rho_{\rm light}(u,\textswab{w})$, given
in (\protect{\ref{eq42}}), at which the
string is moving at the local velocity of light. All the string solutions
cross this curve, with those with smaller (larger) values of $\Pi$
crossing it at smaller (larger) values of $u$.
}
\label{fig1} }

We have solved the radial equation of motion (\ref{eq16}) for $\rho(u)$.
In Fig.~\ref{fig1},
we show results for $\rho(u)$ for different combinations of
the angular velocity
$\textswab{w}$
and the constant of the motion $\Pi$.
It proves convenient to use
$(u_c, \rho_c)$ as the initial point for the numerical evaluation.
For each $\Pi$, the
equations (\ref{eq43}) and (\ref{eq44}) tell us $u_c$
and $\rho_c=\rho(u_c)$.  And, as we saw in Section 2.2, the
equations of motion then determine
$\rho'(u_c)$.
Knowing $\rho(u_c)$ and
$\rho'(u_c)$, we can solve the
second order differential equation (\ref{eq16}) for $\rho(u)$
upward and downward from $u_c$, obtaining the solutions
plotted
in Fig.~\ref{fig1}.  Following the solution upwards
to large $u$ determines what value of $\ell=\rho(\infty)$ corresponds
to the value of $\Pi$ whose selection specified the
particular solution just constructed. From 
the perspective of the rotating quark at the
boundary, the configuration is naturally specified by giving $\ell$
and $\textswab{w}$.  From the perspective of the string to
which this quark is attached, the configuration is naturally
specified by $\Pi$ and $\textswab{w}$.  Fig.~\ref{fig1} illustrates
the fact that there is a one-to-one, and in fact monotonic,
relationship between $\ell$ and $\Pi$.  Following the solution
downwards, we find that at the horizon $\rho(1)$ is finite; we shall
see in the next subsection that there are nevertheless infinitely many
coils of the spiral as the string approaches the horizon.

On the D3-brane at $u=\infty$,
the rotating quark
must not exceed the velocity of light. This limits the radius $\ell$
to $\ell < 1/\textswab{w}$. Indeed, we see in Fig.~\ref{fig1}
that increasing $\Pi$ without bound corresponds to increasing the $u_c$
at which the velocity of the string exceeds the local velocity of light
at that depth in the bulk,
but it never causes $\ell\textswab{w}$ to exceed $1$.
Within the bulk, the string worldsheet
bends towards larger values of $\rho(u)$,
as the string approaches the horizon.
This bending results from the centrifugal force
which the rotating string experiences. It
is more pronounced if the quark moves with greater
acceleration $\textswab{a} = \textswab{w}^2\ell$.
Indeed, we see from the figures
that for a given $\textswab{w}$, there is more bending for
solutions with larger $\ell$. And, the increased bending with
increasing $\textswab{w}$ is also manifest.

We see in Fig.~\ref{fig1} that for $\textswab{w}=0.05$ and $0.5$ the
solutions $\rho(u)$ change in character with increasing $\Pi$: for
small enough $\Pi$, $\rho(1)\simeq \ell$ whereas for large enough $\Pi$
the string bends outward sufficiently that $\rho(1)$ becomes significantly
larger than $\ell$.  However, for $\textswab{w}=5.0$ we find significant
bending for any nonzero $\Pi$.  We have checked these observations
against more solutions than are shown in the figures themselves.  We
find that for $\textswab{w}>1$ there seems to be no range of values of $\Pi$
for which $\rho(1)\simeq \ell$.   We will see in Section 3 that this
crossover from string worldsheets that hang down at almost constant $\rho$
to those that bend significantly outward is associated with a crossover
in the parametric dependence of $dE/dt$, the energy that the rotating
quark loses to the plasma that it is stirring.  We have also investigated
how $\rho(1)$ changes with parameters in the regime in which it is $\gg \ell$.
We find that $\rho(1)$ tends to a finite limit if we increase $\textswab{w}$ while
holding $v$ fixed, while $\rho(1)$ increases without bound if we take $v\rightarrow 1$
while holding $\textswab{w}$ fixed.

For use in  Section 3, we close this subsection by describing the behavior
of $\rho(u)$ in two limits:
\begin{enumerate}
	
\item \underline{The limit $v = \ell\textswab{w} =$ constant,
$\textswab{a} = v\textswab{w} \to 0$, meaning $\textswab{w}\to 0$.}\\
In this limit, we can expand $\rho(u)$ in (\ref{rhoEOM})
as a power series in $\textswab{w}^2$.
Because we are taking $\textswab{w}\to 0$ at fixed $v$, 
we must take $\ell\propto1/\textswab{w}\to\infty$.  The expansion
of $\rho$ therefore starts at order $1/\textswab{w}$, and so can
be written as an expansion of $v(u) \equiv \textswab{w}\rho(u) $:
\begin{equation}
v(u) = v_0 (u) + \textswab{w}^2 v_1 (u) + \ldots
\end{equation}
To leading order, we then find from (\ref{rhoEOM}) that $v_{0}'(u)=0$,
meaning that $v_0(u)={\rm constant}=\ell\textswab{w}=v$. In this limit,
$\rho(u)$ is constant, meaning it is vertical 
in Fig.~\ref{fig1}, and 
its value diverges proportional to $1/\textswab{w}$.
We see from (\ref{eq44}) that,
because $\rho(u)$ is constant and so $\rho_c\textswab{w}=v$,
in this limit  $\Pi\omega\to{\rm constant}$.

\item \underline{The limit $v = \ell\textswab{w} =$ constant,
$\textswab{a} = v\textswab{w} \to \infty$, meaning $\textswab{w}\to \infty$.}\\
In this limit,
$\ell\propto 1/\textswab{w}\to 0$. 
We find that in this limit $\rho(1)\to{\rm constant}$
even though $\ell\to 0$ and, from (\ref{eq44}), $\rho_c\to 0$, meaning that the ratio
$\rho(1)/\ell\to\infty$. We shall discuss this limit further in Section 2.5, where
we shall show that in this limit the string worldsheet takes on the same
shape as in vacuum everywhere except near the horizon, and
where we shall see that in
this  limit $\Pi/\omega\to{\rm constant}$.

\end{enumerate}

%%%%%%%%%%%%%%%%%%%%%%%%%%%%%%%%%%%%%%%%%%%%%%%%
\subsection{Angular dependence of the spiralling string}

\FIGURE[t]{
\includegraphics[width=14cm,angle=270]{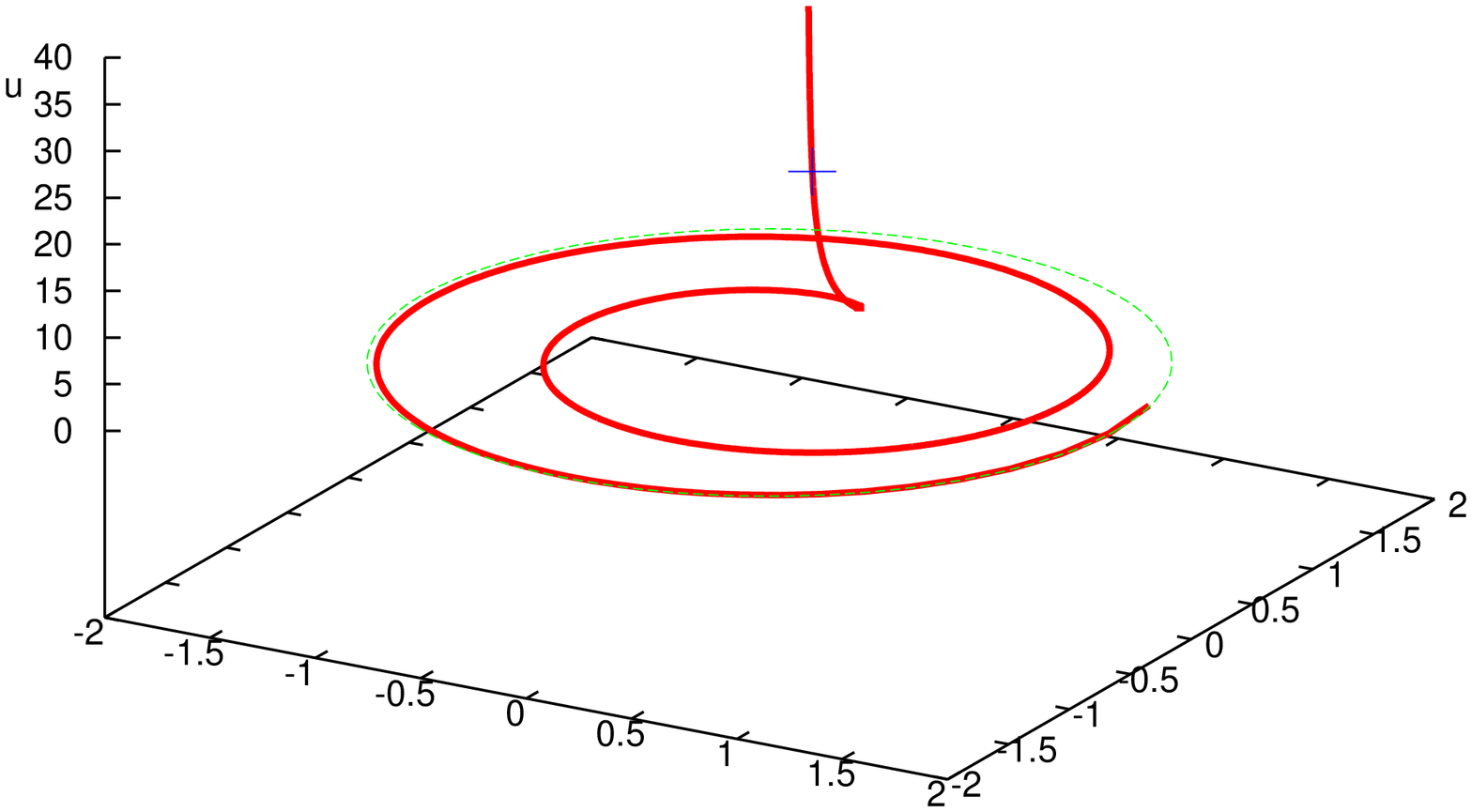}
\caption{The rotating string solution for $\omega = 5.$
and $\Pi = 100.$. The quark rotates clockwise at the boundary at a
radius $\ell=\rho(\infty) = 0.1789$. This
corresponds to a relativistic quark moving with speed $v=0.895$.
The cross marks the depth at which the velocity of the string exceeds
the local velocity of light.
We have ended the plot at $u=1.005$, but the solution actually
spirals downward forever, with infinitely many coils getting
ever closer to the horizon at $u=1$.  The radii of these coils
tend to $\rho(1)=1.65004$, significantly greater than $\ell$
but finite.  The dashed (green) circle denotes
$\rho=\rho(1)$ at $u=1$.
}
\label{fig2} }

Having determined $\rho(u)$ for a fixed angular velocity $\textswab{w}$
and constant of motion
$\Pi$, we can proceed to integrate $\theta'$ given by (\ref{eq19}). This
equation determines $\theta(u)$ up to an overall sign.
The choice of this
sign amounts to deciding whether the quark rotates clockwise or
counter-clockwise in the
$(x_1,x_2)$-plane. We choose the sign positive,
which corresponds to clockwise rotation.
In Fig.~\ref{fig2}, we show a typical example of a
complete solution for $\rho(u)$ and $\theta(u)$, describing
a spiralling string dragging behind and below a rotating quark.
We see the string dropping down from the boundary and spiralling behind
the rotating quark.  The entire spiralling string in
Fig.~\ref{fig2} is rotating with constant angular velocity.
As $u$ decreases, $\rho(u)$ increases as we have already seen
in Fig.~\ref{fig1}. The angle
\begin{equation}
	\theta(u) = \int_u^\infty \theta'(\tilde{u})\, d\tilde{u}
	\label{eq55}
\end{equation}
increases. (Recall that $\theta'$ is given by (\ref{eq19}).)
For $u \to 1$, the curve approaches the horizon at a finite
finite radius $\rho(1)$, with $\rho'(1)$ also finite.
In approaching $u\to 1$,
the string coils around the center of motion of the quark infinitely
many times.  This can be seen from (\ref{eq19}) by noting that
as $u\to 1$, $\theta'$ has a singularity $\propto 1/(u-1)$.
Consequently, there
is a logarithmic increase in $\theta(u)$ as $u\rightarrow 1$.
Explicitly, if we define
$\Delta\theta(u)$ as the angle swept out by the string as it
descends from some reference value of $u$, which
we denote $u_0$, that is already close to the horizon, $u_0-1\ll 1$,
down to values of $u$ that are even closer to the horizon,
then
\begin{equation}
	\Delta\theta(u)  = \int_u^{u_0} \theta'(\tilde{u}) d\tilde{u}
		\simeq \int_u^{u_0}  \frac{\textswab{w}}
{\tilde{u}^4 - 1}
d\tilde{u}
\simeq \int_u^{u_0}  \frac{\textswab{w}}
{4(\tilde{u} - 1)}
		=
\frac{\textswab{w}}{4}
\ln \left[ \frac{u_0-1}{u-1}\right]\, .
		\label{eq56}
\end{equation}
This behavior of $\theta(u)$ near the horizon
is quite analogous to the behavior of the trailing string
hanging from a quark moving with constant linear velocity, $x_1=vt$.
This string worldsheet is given by~\cite{Herzog:2006gh,Gubser:2006bz}
\begin{equation}
x_1(u,t) = v t - \frac{v}{2}
\left[\frac{\pi}{2}-\arctan u - {\rm arccoth}\, u \right]\ ,
\label{linearstring}
\end{equation}
with both $t$ and $x_1$ measured in units of $1/(\pi T)$.
This means that
\begin{equation}
x_1' =  \frac{v}{u^4-1} \simeq \frac{v}{4(u-1)}
\label{linearx1prime}
\end{equation}
in the vicinity of the horizon. The logarithmic divergence
of the length of this trailing straight string and of the length
of our trailing rotating string are clearly analogous.
They are not the same because the
quantity $\rho(1)\textswab{w}$ which appears in (\ref{eq56})
in the same way that the linear velocity $v$ appears
in (\ref{linearx1prime}) is in fact greater, and in some
instances much greater, than the speed
of the rotating quark, which is given by $\ell\textswab{w}$.

\subsection{Rotating quark in vacuum} \label{sec:vacuum}

In the next Section, we shall want to compare the
energy lost by a rotating quark in the hot
plasma of strongly coupled ${\cal N}=4$ SYM theory to that
lost by a rotating quark in vacuum in the same theory.
To that end, it will be useful to understand the
ways in which the trailing string in vacuum is similar to or different
from the trailing string in the plasma that we have constructed.
In vacuum, the string describes the energy radiated from
the rotating quark due to its acceleration.  The calculation
in vacuum proceeds analogously to that above, and we
therefore need only sketch it briefly.

In vacuum, the Lagrangian (\ref{eq15}) and the equations of
motion (\ref{eq19}) and (\ref{rhoEOM})
simplify to
\begin{equation}
	{\cal L} =
	\sqrt{ \left(1 - \rho^2 \textswab{w}^2 \right)
		\left(u^4\rho'^2 + 1\right) +
	         \rho^2 u^4 \theta'^2}
	         \label{vacuumLag}
\end{equation}
and
\begin{equation}
	{\theta'}^2 = \frac{\Pi^2\, \left(1-\rho^2\textswab{w}^2 \right) \left(u^4 \rho'^2
	+ 1\right)}{\rho^2 u^4 \left(\rho^2 u^4 - \Pi^2\right)}\, ,
\label{vacuumthetaEOM}
\end{equation}
and
\begin{equation}
\rho''+\frac{2\rho'}{u} + \left(1+\rho'^2u^4\right)\rho \left( \frac{2u^3 \rho\rho'-1}{u^4\rho^2-\Pi^2}
+\frac{1}{\rho^2u^4(1-\rho^2\textswab{w}^2)}\right)=0\ ,
\label{vacuumrhoEOM}
\end{equation}
where $\Pi=\partial{\cal L}/\partial \theta'$ is a constant of the motion as before
and where the dimensionless variables $u$,  $\rho$, 
and $\textswab{w}$ are now related
to their dimensionful counterparts by $u =  \tilde u/R$, $ \rho = \tilde\rho/R$,
and $\textswab{w}=\omega R$
instead of (\ref{eq12}),  (\ref{eq13}) and (\ref{eq14}).

As before, the denominator of (\ref{vacuumthetaEOM}) vanishes
at some $u_c$ and becomes negative for $u<u_c$, meaning that
the numerator of (\ref{vacuumthetaEOM}) must vanish at the same $u_c$.
In vacuum, this worldsheet horizon where the string velocity equals the
local speed of light is located at
\begin{eqnarray}
u_c^{\rm vac}&=&\sqrt{\Pi \textswab{w} }\label{vacuum_uc}\, ,\\
\rho_c^{\rm vac}&=&\frac{1}{\textswab{w}}\label{vacuum_rhoc}\ .
\end{eqnarray}

As before, choosing $\Pi$ determines $(u_c^{\rm vac},\rho_c^{\rm vac})$
and the equation of motion (\ref{vacuumrhoEOM}) can then be solved starting
from this point.  Following the solution upwards determines $\ell$, 
where $\ell R$ is the radius
of the circle at $u=\infty$ along which the quark is moving.
Notice that in the large $\Pi\textswab{w}$ limit, the location of the
world sheet horizon for a quark rotating in plasma
given by (\ref{eq43}) and (\ref{eq44}) tends
to $(u_c^{\rm vac},\rho_c^{\rm vac})$.  Correspondingly, we find that if
we choose $\Pi\textswab{w}$ to be large,
the solution $\rho(u)$ that
we obtain by solving (\ref{vacuumrhoEOM})
for $u>u_c^{\rm vac}$ is very similar to that which we found previously
in plasma for $u>u_c$.

Following the solution $\rho(u)$ downwards from $(u_c^{\rm vac},\rho_c^{\rm vac})$,
we find that in this regime the string worldsheet is qualitatively different
in vacuum relative to that which we found previously in the plasma.  Recall that
in the plasma the solution extends down toward the horizon at $u=1$, where $\rho(1)$
is finite and where $\theta(u)$ increases logarithmically as $u\rightarrow 1$.
Instead, in vacuum the solution extends down toward $u=0$ and it is easy
to show from (\ref{vacuumrhoEOM}) and (\ref{vacuumthetaEOM}) that
as $u\rightarrow 0$ the solution takes the form
\begin{eqnarray}
\rho(u)&\approx& \frac{f}{u}+\ldots\label{rhovacsmallu}\\
\theta(u)&\approx&\frac{\sqrt{1+f^2}}{u}+\ldots\label{thetavacsmallu}
\end{eqnarray}
for some constant $f$, meaning that the spiralling
string world sheet extends to $\rho\rightarrow\infty$ as $u\rightarrow 0$.
We shall discuss the interpretation of this qualitative distinction between
the pattern of energy deposited by a quark rotating in vacuum and one
rotating in the plasma in Section 3.2.

It is instructive to notice that if we
rescale $u$ and $\rho$ by introducing
new variables $u/\textswab{w}$ and $\textswab{w}\rho $ then,
when written in terms of the new variables, both the equations of motion
(\ref{vacuumthetaEOM})
and (\ref{vacuumrhoEOM}) include $\Pi$ and $\textswab{w}$ only in the
combination $\Pi/\textswab{w}$. This means
that there is a one-to-one map between the velocity of the quark $v$ and 
the ratio $\Pi/\textswab{w}$.
Even though this simplification
of the vacuum equations does not occur
in the  $T\neq 0$ equations, it prompts us to return to the equation
of motion (\ref{rhoEOM}) for $\rho(u)$ at $T\neq 0$ and reconsider the
limit in which we take $\textswab{w}\to\infty$ at fixed $v$,
introduced briefly in Section 2.3.
If we rewrite (\ref{rhoEOM}) in terms of $u/\textswab{w}$ and $\textswab{w}\rho$ 
it still contains
both $\Pi/\textswab{w}$ and $\textswab{w}$, but if we
then take the $\textswab{w}\to\infty$ limit at fixed $\Pi/\textswab{w}$
by expanding the equation to zeroth order
in powers of $1/\textswab{w}$ we find
that we recover the zero temperature equation (\ref{vacuumrhoEOM}),
with (\ref{eq43}) and (\ref{eq44}) becoming (\ref{vacuum_uc}) and (\ref{vacuum_rhoc})
as we already saw.  Thus, in this 
scaling limit the nonzero temperature and 
zero temperature solutions $\rho(u)$  become identical for $u\geq u_c$
and the relation between $\Pi/\textswab{w}$ and the
quark velocity $v$ becomes the same as in vacuum. 
However, this analysis breaks down near the black hole horizon.
At the horizon, 
$\rho(u)\to\rho(1)={\rm constant}$
in the $\textswab{w}\to\infty$ at fixed $v$ limit, 
meaning that $\textswab{w}\rho(1)\propto \textswab{w}$.
This rearranges the $1/\textswab{w}$-expansion of (\ref{rhoEOM})
in the vicinity of the horizon, since what must now
be expanded is the equation written 
in terms of $\rho$ rather than $\textswab{w}\rho$.
The $T\neq 0$ equation of motion then no longer reduces to its vacuum form.
Our numerical solutions confirm that for all $u>u_c$ and for much of
the $u<u_c$ region (but not near $u=1$) as we increase $\textswab{w}$
at fixed $\Pi/\textswab{w}$, and hence at fixed $v$, the solution
$\rho(u)\propto 1/\textswab{w} \to 0$ and approaches its shape in vacuum.
Near the horizon, however, where (\ref{rhoEOM}) does not become (\ref{vacuumrhoEOM}), 
we have $\rho(u)\to\rho(1)={\rm constant}$.  This difference between
the behavior of $\rho(u)$ near the horizon and its behavior in vacuum
is responsible for the difference between (\ref{rhovacsmallu}) and (\ref{thetavacsmallu})
and the behavior of $\rho$ and $\theta$ near the horizon at $T\neq 0$, described
in Sections 2.3 and 2.4.

%%%%%%%%%%%%%%%%%%%%%%%%%%%%%%%%%%%%%%%%%%%
\section{Energy Lost by a Rotating Quark}
\label{sec3}

\subsection{General and numerical results}

As Herzog {\it et al} and Gubser showed for the case of a quark moving
in a straight line with constant speed~\cite{Herzog:2006gh,Gubser:2006bz},
in a setting in which a moving quark in the boundary theory is
described by a string worldsheet whose shape does not change with
time, the energy lost by the moving quark is easy to evaluate once
the shape of the string worldsheet has been determined.
The energy deposited in
the medium per unit time (or, equivalently, the
power expended by the external agent moving the quark) is given by
\begin{equation}
\frac{dE}{dt} = \Pi_t^\sigma \ ,
\end{equation}
where
\begin{equation}
\Pi_\mu^\sigma \equiv \frac{1}{2\pi\alpha'}\,
\frac{\partial {\cal L}}{\partial \left(\partial_\sigma X^\mu\right)}
= - G_{\mu\nu} \frac{ \left[(\partial_uX)\cdot(\partial_\tau X) \right]
\partial_\tau X^\nu
	     - \left[ (\partial_\tau X)\cdot(\partial_\tau X) \right]
\partial_uX^\nu}
	{2\pi\alpha'\sqrt{-{\rm det} g_{ab}} }
\label{PiDefn}
\end{equation}
gives the flow of either energy ($\mu=t$) or momentum down the string.
For the rotating string that we have analyzed in
Sections 2.1-2.4, explicit evaluation yields a result with the simple form
\begin{equation}
	\frac{dE}{dt}
		= \frac{\pi}{2} \sqrt{\lambda} T^2  \textswab{w} \Pi \, ,
			\label{eq22}
\end{equation}
where we have restored all the dimensionful factors and have
used (\ref{LambdaAlpha}).\footnote{
We have checked that the same result can also be obtained by evaluating
\begin{equation}
\frac{dE}{dt} = - \frac{d}{dt} \int du \, \Pi_t^\tau
\end{equation}
(where $\Pi_t^\tau$, which can be obtained by interchanging $\tau$
and $\sigma$ in (\ref{PiDefn}), is the energy density $dE/du$ along the string)
upon choosing any time independent upper cutoff to the $u$-integral
and upon choosing a (time-dependent)
lower cutoff to the $u$-integral corresponding to
the $u$ of the string worldsheet at some fixed angle $\theta$.}

It will prove instructive to recast our result (\ref{eq22}) for the energy loss in terms
of the location $(u_c,\rho_c)$ of the point at which the velocity of the
rotating string exceeds the local velocity of light at that depth in the bulk
or, equivalently, the location of the worldsheet horizon.
By solving (\ref{eq44}) for $\Pi\textswab{w}$, we find
that $dE/dt$ in (\ref{eq22}) is given by
\begin{equation}
	\frac{dE}{dt}
		= \frac{\pi}{2} \sqrt{\lambda} T^2 \frac{v_c^2}{\sqrt{1-v_c^2}}\ ,
\label{eq22c}
\end{equation}
where $v_c\equiv \textswab{w} \rho_c$ is the velocity of the string at the
point where that velocity is also the local velocity of light.
We can compare our result (\ref{eq22c})
to $dE/dt$ for a quark moving
in a straight line with speed $v$, which is easily obtained
from the string worldsheet (\ref{linearstring})
and is given by~\cite{Herzog:2006gh,Gubser:2006bz}
\begin{equation}
\frac{dE}{dt}\Biggl|_{\rm linear\ drag} = \frac{\pi}{2}\sqrt{\lambda}\,T^2
\frac{v^2}{\sqrt{1-v^2}}\ .
\label{lineardrag}
\end{equation}  	
Because the general result (\ref{eq22})
for the energy loss of a rotating quark
takes the form (\ref{eq22c}), it is
identical to that for a quark
in linear motion with velocity $v_c$. Wherever $v_c\simeq v$, the
standard linear drag result for $dE/dt$ is obtained.  Wherever the
outward curvature of the string worldsheet in Fig.~\ref{fig1}
is significant, making $v_c$ significantly greater than $v$, $dE/dt$ is
greater. We shall interpret
this result further
in Section 3.2.\footnote{In order to make contact
with Ref.~\cite{Dominguez:2008vd}, we can
instead solve (\ref{eq43})
for $\Pi\textswab{w}$,  finding that $dE/dt$ in (\ref{eq22}) is given by
\begin{equation}
	\frac{dE}{dt}
		= \frac{\pi}{2} \sqrt{\lambda} T^2 \frac{u_c^4-1}{u_c^2}\ .
\label{eq22a}
\end{equation}		
In the $v=\ell\textswab{w}\rightarrow 1$ limit, where the worldsheet
horizon $u_c\rightarrow \infty$, we see from (\ref{eq43}) that
$u_c\simeq \sqrt{\textswab{w} \Pi}$, yielding the even simpler
result
\begin{equation}
	\frac{dE}{dt}
		= \frac{\pi}{2} \sqrt{\lambda} T^2  u_c^2 \, .
			\label{eq22b}
\end{equation}
At high velocity
we therefore reproduce the relationship
between $dE/dt$ and the worldsheet
horizon for a quark moving in a straight
line at constant velocity in the high velocity limit
that has been emphasized in
Ref.~\cite{Dominguez:2008vd}.}

The result (\ref{eq22}) makes it straightforward to
use the numerical calculations described in Section 2 to evaluate
the energy loss $dE/dt$ for a quark moving in a circle with
angular frequency $\omega=\textswab{w}\pi T$ as a function of
the radius of the circle  $L=\ell/(\pi T)$, as follows. We pick
$\textswab{w}$ and we pick a series of values of $\Pi$. For each $\Pi$,
we follow the procedure described in Section 2 to obtain the corresponding
string worldsheet. From that solution, we determine $\ell$ at large $u$.
Then, since $dE/dt$ is just a constant times $\Pi$,
we plot $\Pi$ as a function of $\ell$, obtaining the plots in the
first row of Fig.~\ref{fig3}.  We see that $\Pi$, and hence $dE/dt$,
increase without bound as $\ell$ approaches $1/\textswab{w}$, which
is the radius at which the quark would be moving at the speed of light.

%%%%%%%%%%%%%%%%%%%%%%%%%%%%%%%%%%%%%%%%%
\FIGURE[t]{
\includegraphics[width=17cm,angle=0]{pvll.eps}
\caption{Upper row: $\Pi$, which is proportional to the
rate of energy loss $dE/dt$ according to (\protect{\ref{eq22}}),
plotted as a function of the dimensionless radius $\ell$ at which the
quark rotates, for three different values of the dimensionless
angular velocity $\textswab{w}$. Second row:
$\Pi/\Pi_{\rm linear\ drag}$, where $\Pi_{\rm linear\ drag}$
is given in (\ref{eq47}) and is what we expect to find if $dE/dt$ is
given by that due to the drag on a quark moving in a straight
line with velocity $v=\ell\textswab{w}$.  We see that at $\textswab{w}=0.05$
and $0.5$, $\Pi\simeq\Pi_{\rm linear\ drag}$ at low enough $\ell$, i.e. at
low enough acceleration $\textswab{a}$.  At $\textswab{w}=5.0$, however,
$\Pi$ exceeds $\Pi_{\rm linear\ drag}$ by a large factor at any $\ell$, e.g. by
a factor of 24.980 for $\ell\rightarrow 0$ and
hence $\textswab{a}\rightarrow 0$.
Third row:  $\Pi/\Pi_{\rm vacuum\ radiation}$, where
$\Pi_{\rm vacuum\  radiation}$ is given
in (\ref{eq48}) and is what we would find if $dE/dt$
is given by the energy loss due to the radiation of an accelerating
charged particle in vacuum.  We see that $\Pi\simeq\Pi_{\rm vacuum\ radiation}$
at large enough $\ell$ for $\textswab{w}=0.05$ and $0.5$  and
at all $\ell$ (and hence all $v$ and $\textswab{a}$) for $\textswab{w}=5.0$.
Bottom row: The ratio $\Pi/\left( \Pi_{\rm vacuum\ radiation} + \Pi_{\rm linear\ drag}\right)$
is smaller than one for all $\omega$ and all $\ell$, supporting the picture that there is
destructive interference between acceleration-induced radiation and medium-induced radiation.
}
\label{fig3} }
%%%%%%%%%%%%%%%%%%%%%%%%%%%%%%%%%%%%%%%%

\subsection{Energy loss in two limits: linear drag and vacuum radiation}

Numerical results for $dE/dt$, as in the first row of Fig.~\ref{fig3},
are not very informative until they are compared to analytic
expectations or explanations. There are two limits to consider:

\begin{enumerate}
	
\item \underline{The limit $v = \ell\textswab{w} =$ constant,
$\textswab{a} = v\textswab{w} \to 0$.}\\
This limit is reached by increasing $\ell$ and decreasing $\textswab{w}$
while keeping $v$ fixed.
Because we are taking the (dimensionless)
acceleration to zero, in this limit we expect $dE/dt$ to be given by
the result (\ref{lineardrag}) for a quark moving in a straight line with speed $v$,
meaning that we expect that as $\textswab{a}\to 0$ at fixed $v$
we should find that $dE/dt$ for the rotating quark is given by (\ref{eq22})
with
\begin{equation}
	\textswab{w} \Pi \to \textswab{w} \Pi_{\rm linear\ drag} \equiv
\frac{v^2}{\sqrt{1-v^2}}\, ,
	\label{eq47}
\end{equation}
%This can be shown analytically
meaning that holding $v$ fixed as we take the $\textswab{w}\to 0$
limit means holding $\Pi\textswab{w}$ fixed.
The general result (\ref{eq22c}) for $dE/dt$ together with the result
from Section 2.3 that
$\rho(u)=\ell={\rm constant}$ and therefore $v_c=v$ in the present 
limit constitute an analytic demonstration of (\ref{eq47}).
In the second row of Fig.~\ref{fig3}, we show the ratio of our
numerically obtained results for $\Pi$ for the rotating quark
to $\Pi_{\rm linear\ drag}$. We see that our $dE/dt$ is described well
by that for a quark in linear motion when $\ell$ is small enough
at $\textswab{w}=0.05$ and $\textswab{w}=0.5$, i.e. in the regime
where $\textswab{a}$ is small.  Surprisingly, we see in the
panel with $\textswab{w}=5.0$ that once the dimensionless frequency
is large enough, $\Pi \gg \Pi_{\rm linear\ drag}$ even at small $\ell$,
where the dimensionless acceleration $\textswab{a}=\ell\textswab{w}^2$
is small.
So, the criterion for the validity of the linear drag approximation
to our result for the energy loss of a rotating quark cannot
be simply $\textswab{a} \rightarrow 0$.

Given the general result (\ref{eq22c}), the validity of the linear drag
approximation can be related directly to the shape of the string profile
$\rho(u)$, as in Fig.~\ref{fig1}.  Whenever $\rho(u)$ is to a good approximation
constant, meaning that the curves in Fig.~\ref{fig1} are to a good approximation
vertical, $\rho_c \simeq \ell$ and hence $v_c \simeq v$, meaning that in this
circumstance the general result (\ref{eq22c}) becomes (\ref{eq47}), as for
a quark in linear motion.

\item
\underline{The limit $v = \ell\textswab{w} =$ constant,
$\textswab{a} = v\textswab{w} \rightarrow \infty$.}\\
This limit is reached by decreasing $\ell$ and increasing $\textswab{w}$
while keeping $v$ fixed.  In this limit, we expect the effects of
acceleration to dominate.  And, we have seen in Sections 2.3 and 2.5 that
in this limit the shape of the spiralling string at $u>u_c$ --- above
the worldsheet horizon --- is the same at $T\neq 0$ as it is in vacuum.
This means that we expect that $dE/dt$ for the quark stirring the
strongly coupled plasma in this acceleration-dominated regime should
be the same as if the quark were rotating in vacuum.
And, in Ref.~\cite{Mikhailov:2003er}
Mikhailov
has derived an elegant and general
result for $dE/dt$ for an accelerating quark in ${\cal N}=4$ SYM theory
at $T=0$, i.e. in vacuum,
where $dE/dt$ is due to (synchrotron) radiation.  Mikhailov's result
is
\begin{equation}
\frac{dE}{dt}\Biggl|_{\rm vacuum\ radiation}
= \frac{\sqrt{\lambda}}{2\pi} \,
\frac{ \vec{a}^2 -
\left(\vec{a}\times\vec{v}\right)^2}{\left(1-v^2\right)^3}\ .
\label{mikhailovresult}
\end{equation}
This result is equivalent to Li\'enard's result for electromagnetic
radiation from an accelerating charge~\cite{Lienard}
upon replacing $2e^2/3$ in
the latter by $\sqrt{\lambda}/(2\pi)$.  For the case of circular motion,
Mikhailov's result becomes
\begin{equation}
\frac{dE}{dt}\Biggl|_{\rm vacuum\ radiation}
= \frac{\sqrt{\lambda}}{2\pi}\, \frac{a^2}{\left(1-v^2\right)^2}
\equiv  \frac{\sqrt{\lambda}}{2\pi}\, a_{\rm proper}^2\ ,
	\label{a7}
\end{equation}
where we have defined the proper acceleration in the usual
fashion.
So, if in the $\textswab{a}\rightarrow\infty$ at fixed $v$ limit
$dE/dt$ for our rotating quark is as it would be due
to radiation in vacuum, we should find that $dE/dt$ is given by (\ref{eq22})
with
\begin{equation}
\pi^2 T^2 \textswab{w} \Pi \to a_{\rm proper}^2 \ , 
\end{equation}
namely
\begin{equation}
\textswab{w} \Pi \to \textswab{w} \Pi_{\rm vacuum\ radiation}
\equiv  \textswab{a}_{\rm proper}^2  \equiv
\frac{\textswab{a}^2}{\left(1-v^2\right)^2}
=\frac{v^2\textswab{w}^2}{\left(1-v^2\right)^2}\ ,
\label{eq48}
\end{equation}
meaning that holding $v$ fixed as we take the $\textswab{w}\to \infty$
limit means holding $\Pi/\textswab{w}$ fixed.
In the third row of Fig.~\ref{fig3}, we show the ratio of our
numerically obtained results for $\Pi$ for the rotating quark
to $\Pi_{\rm vacuum\ radiation}$. We see that our $dE/dt$ is described well
by that for a quark in circular motion that is
radiating in vacuum when $\ell$ is close enough to $1/\textswab{w}$
at any $\textswab{w}$.  Thus, in the large $\textswab{a}$ limit
the energy loss of a quark that is stirring
the plasma is the same as it would be due
to its acceleration if it were rotating in vacuum.  Furthermore,
the $\textswab{w}=5.0$ panel shows that at large enough $\textswab{w}$
this result extends to all $\ell$, including down to
values of $\ell$ that are small enough that $\textswab{a}$ is
small.
So, the criterion for the validity of the vacuum radiation approximation
to our result for the energy loss of the rotating quark cannot
be simply $\textswab{a} \rightarrow \infty$.
\end{enumerate}

It turns out that the correct criterion for determining under what
circumstances $dE/dt$ is as it would be due to the drag on a quark
moving in a straight line with the same $v$ and under what
circumstances it is as it would be due to radiation if the quark  were
rotating in vacuum 
is given simply by asking which of the two yields the larger $dE/dt$.
$dE/dt|_{\rm vacuum\ radiation} \gg dE/dt|_{\rm linear\ drag}$ when
$\Pi_{\rm vacuum\ radiation} \gg \Pi_{\rm linear\ drag}$, namely  when
\begin{equation}
\textswab{w} \gg \left(1-v^2\right)^{3/4}\ ,
\label{eq51}
\end{equation}
where we have used (\ref{eq47}) and (\ref{eq48}).
We see in particular that
energy loss due to radiation in vacuum is larger than that due
to linear drag in a plasma when $\textswab{w}>1$ for any $v$.
This explains the right panels of Fig.~\ref{fig3}.  In the middle
and left panels of Fig.~\ref{fig3}, the location of the crossover between
$dE/dt \simeq dE/dt|_{\rm linear\ drag}$ at lower $\ell$ and
$dE/dt\simeq dE/dt|_{\rm vacuum\ radiation}$ at higher $\ell$ is well described
by the criterion (\ref{eq51}).  Furthermore, we find that the
regime where
the $\gg$ in (\ref{eq51}) is a $\sim$ is the same regime where,
in Section 2, we saw the radial profile $\rho(u)$ of the
spiralling string hanging down from the rotating quark begin
to bend outward significantly.  Where the $\gg$ in (\ref{eq51})
is a $\ll$, the $\rho(u)$ curves in Fig.~\ref{fig1} are vertical to a
very good approximation.

If we plot $\Pi/(\Pi_{\rm linear\ drag}+\Pi_{\rm vacuum\ radiation})$ as
in the fourth row of Fig.~\ref{fig3}, we find
that $\Pi\simeq (\Pi_{\rm linear\ drag}+\Pi_{\rm vacuum\ radiation})$ in the regimes
where one or other dominates, and in the crossover between the drag-dominated
regime and the acceleration-radiation-dominated regime we find
$\Pi<(\Pi_{\rm linear\ drag}+\Pi_{\rm vacuum\ radiation})$, as if the two
sources of energy loss were interfering destructively.
For sufficiently small $\omega$, the minimum in the ratio
$\Pi/(\Pi_{\rm linear\ drag}+\Pi_{\rm vacuum\ radiation})$ is at about 2/3, and
this minimum
occurs close to the point
at which the inequality (\ref{eq51}) turns into an equality, namely
$\ell = \sqrt{1-\omega^{4/3}}/\omega$.

\FIGURE[t]{
\includegraphics[width=17cm,angle=0]{xxvsyy4.eps}
\caption{Scaled energy loss versus scaled velocity.  By plotting
$\Pi\textswab{w}/(\gamma v^2)$ versus $\textswab{w}^2\gamma^3$, we can
illustrate the crossover between the regime in which energy
loss is dominated by drag --- where  $\Pi\textswab{w}/(\gamma v^2)=1$ ---
and the regime in which energy loss is dominated by
radiation as if in vacuum --- where
 $\Pi\textswab{w}/(\gamma v^2)=\textswab{w}^2\gamma^3$.
Furthermore, using these scaled variables our results
for any value of $\textswab{w}$ come close to falling on
a single  curve.  For each of the three values  of $\textswab{w}$
plotted, the star marks the point at which $v=0$ and $\gamma=1$ and
the curve extends from the star arbitrarily far to the right as $\gamma$ increases.
For any $\textswab{w}$,  the
regime where radiation due to acceleration dominates the energy
loss is reached at large enough $\gamma$; for $\textswab{w}=5$, the energy
loss is in this regime already at $\gamma=1$.  We have also plotted
$\Pi\textswab{w}/(\gamma v^2)=1+\textswab{w}^2\gamma^3$, which is
the result that we would have obtained if the energy loss  were the incoherent
sum of that due to drag and that due to radiation as if in vacuum. Our results
lie below this curve, indicating destructive interference.
}
\label{fig4}}

The interplay between energy loss due to linear drag and
energy loss due to acceleration-induced radiation as if in vacuum
can be illustrated in a single figure by plotting
$\Pi\textswab{w}/(\gamma v^2)$ versus $\textswab{w}^2\gamma^3$,
where $\gamma\equiv 1/\sqrt{1-v^2}$.
The resulting Fig.~\ref{fig4}  
illustrates clearly that the energy loss is dominated by
linear drag when $\textswab{w}^2\gamma^3\ll 1$ and by radiation as
if in vacuum when $\textswab{w}^2\gamma^3\gg1$, as in (\ref{eq51}).
We understand from (\ref{eq47}) and (\ref{eq48}) 
why the curves with different values of $\textswab{w}$
lie on top of each other in these regimes, 
but we have no analytical
argument why there should be a single scaling curve in the crossover regime.
We observe, however, that the curves come very close to lying on a
single universal scaling curve even in the crossover regime.
Looking back
at Fig.~\ref{fig3}, we realize from Fig.~\ref{fig4} that the dips in the curves in
the lower left and lower middle plots in  Fig.~\ref{fig3}
would have the same shape if they were plotted versus $\textswab{w}^2\gamma^3$,
rather than versus $\ell$.  In Fig.~\ref{fig4} we have also illustrated
the evidence for destructive interference between energy loss due to radiation
as if in vacuum and energy loss due to drag by showing that our results lie
below where they would have been if the energy loss were simply
the incoherent sum of that due to these two mechanisms.

Our results illustrate that the analysis of the energy loss
of a rotating quark that is stirring the plasma allows us to study
the crossover from a linear-drag-dominated regime to an
acceleration-dominated regime in a calculation that is valid
in both regimes. (In Section 3.3, we shall discuss where our calculation
breaks down.)
In the acceleration-dominated regime, $dE/dt$ is as it would be in
vacuum.  We can give a partial explanation for this phenomenon.
Given that Mikhailov's result for radiation of an accelerating charge in ${\cal N}=4$ SYM
differs only by the coupling constant from Li\'enard's result for QED, it is
natural to expect
that, as in QED, the spectrum of the synchrotron radiation
rises gradually with increasing frequency until frequencies $\simeq \omega_c$,
and then falls off exponentially for frequencies that are $\gg \omega_c$,
with the critical frequency being given by
\begin{equation}
\omega_c = 3 \gamma^3 \omega \ ,
\label{eqq}
\end{equation}
with $\gamma=1/\sqrt{1-v^2}$ and $\omega$ the dimensionful
angular frequency of the rotating charge.
This means that almost all of the radiated
energy is carried by radiation modes with frequencies of order $\omega_c$.
And, in any circumstance in which the criterion
(\ref{eq51}) is satisfied, $\omega_c \gg \pi T$. (To see this,
note that (\ref{eq51}) can either be satisfied via $\textswab{w}\gg 1$,
which means $\omega\gg \pi T$ and hence $\omega_c \gg \pi T$ even
if $\gamma$ is not large, or (\ref{eq51}) can be satisfied at small
$\textswab{w}$ for values of $\gamma$ that are large enough that,
again, $\omega_c \gg \pi T$.)  So, if the plasma were weakly coupled
we could conclude that radiation into modes with
frequencies $\omega_c \gg \pi T$ is unaffected by the presence
of the plasma and thus understand why $dE/dt$ in the acceleration-dominated
regime (\ref{eq51}) is as it would be if the rotating quark were
emitting synchrotron radiation in vacuum.  However, in ${\cal N}=4$ SYM
we have a plasma that is strongly coupled at all scales. And, the synchrotron
radiation is composed of colored excitations rather than neutral photons.
We do not expect that any excitations propagate for times large compared to the inverse
of the temperature --- there are after all no long-lived quasiparticles
expected in this system.  And, we expect any energetic colored excitations
to be strongly quenched by the plasma.

The shape of the spiralling string worldsheet trailing behind the rotating
quark, as in Fig.~\ref{fig2} provides direct evidence that the behavior
of the energy that is deposited in the plasma by the rotating quark
is different from that of synchrotron radiation in vacuum even in the
acceleration-dominated regime in which $dE/dt$ is as for vacuum radiation.
We saw in Section 2.5 that when we redo our calculation in vacuum, the
rotating string extends out to $\rho\rightarrow\infty$. This corresponds to
the fact that, in vacuum, the energy that is radiated propagates out
to infinity.  Instead, in the plasma we find that as  the string
approaches the horizon it coils on top of itself over and over
again wiith $\rho$ tending to a finite $\rho(1)$ at the horizon.
This indicates that the energy deposited in the plasma (which can
be thought of as having been left behind or radiated by the rotating
quark) only spreads outwards to some finite radius.   In the
drag-dominated regime, where $\rho(1)=\ell$, the quark leaves
a wake behind along its trajectory, as for linear motion at
constant velocity.  In the acceleration-dominated regime, where
$\rho(1)>\ell$, the deposited energy spreads  outwards, but not
to infinity.  So, in the acceleration-dominated regime $dE/dt$
is as if the rotating quark were emitting synchrotron radiation
in vacuum, but the behavior of the energy that the quark deposits
in the plasma is different than that of synchrotron radiation in vacuum.

One way to gain further insight
would be to calculate
the stress-energy tensor for the plasma being stirred by the rotating quark.
For the
case of a quark in linear motion, the work of Refs.~\cite{Friess:2006fk}
has illuminated how the strongly coupled plasma carries the
energy lost by the moving quark.  By doing such a calculation
for a rotating quark stirring the strongly coupled plasma,
we could see whether in the acceleration-dominated regime
the stress energy tensor in the region $\rho<\rho(1)$
takes on the characteristic form
expected for synchrotron radiation, with a pulse of energy density
passing any distant point once every $2\pi/\omega$ in time, and
with these pulses having a width $\sim L/\gamma^3$ that narrows
with increasing $v$.   We have seen in Section 2.3
that $\rho(1)$ tends to a constant
if $\omega$ is increased at constant $\gamma$ while $\rho(1)$
increases without bound if $\gamma$ is increased at constant $\omega$.  This
suggests that the distance that the energy left behind by the rotating
quark spreads before it thermalizes is controlled by the width
of the synchrotron radiation pulses, not by the time between pulses.
This speculation
could be tested via a calculation of the stress energy tensor, which
we leave to future work.

%%%%%%%%%%%%%%%%%%%%%%%%%%%%%%%%%%%%%%%%%%%
\subsection{Where the calculation breaks down}

The calculation of $dE/dt$ in (\ref{lineardrag}) for a quark moving
in a straight line with constant velocity is known to break
down at sufficiently large velocity for quarks with finite mass $M$.
We can give our test quarks finite mass
by putting the D3-brane
on which they live at $u=\Lambda$ with $\Lambda$
finite and $\Lambda\gg 1$,
yielding quarks with mass
\begin{equation}
M= \frac{\sqrt{\lambda} T \Lambda}{2} \ .
\label{QuarkMass}
\end{equation}
(This is equivalent, for our purpose, to the more sophisticated
approach~\cite{Karch:2002sh}
of introducing a D7-brane that fills the part of the AdS
space with $u>\Lambda$.)  Then, the calculation of the
trailing string (\ref{linearstring}) and the associated linear
drag (\ref{lineardrag}) breaks down when the worldsheet horizon
located at $u=u^{\rm linear}_c=\left(1-v^2\right)^{-1/4}$
reaches $\Lambda$~\cite{CasalderreySolana:2007qw}.  One way to
see that something must go wrong is to realize that if (\ref{linearstring})
were valid when $v$ is so large that $u_c^{\rm linear}>\Lambda$ then
the speed of the quark itself,  at the D3-brane, would exceed that of light.
Casalderrey-Solana and Teaney provided an equivalent and more physical
understanding of what breaks down  by noting that at this velocity
the force required to keep the quark moving with this velocity becomes
so great that the production of pairs of quarks  and antiquarks with
mass $M$ becomes
unsuppressed~\cite{CasalderreySolana:2007qw}.\footnote{At the same velocity,
the screening length characterizing the potential between a moving quark
and antiquark becomes shorter than the quark Compton
wavelength~\cite{Liu:2006he}.} By the same argument, our calculation
breaks down when $\Pi$ gets so large that $u_c=\Lambda$.
Upon noting from (\ref{eq43}) that at large $\Pi$ we have
$u_c\simeq \sqrt{\Pi\textswab{w}}$, this means that our calculation
breaks down when $\Pi\textswab{w}=\Lambda^2$.

\FIGURE[t]{
\includegraphics[width=17cm,angle=0]{fig5b.eps}
\caption{Regime of validity of our classical worldsheet calculation in
the $(\textswab{w},\gamma)$-plane as a function
of quark mass.  The calculation is
valid for $\Pi\textswab{w}<\Lambda^2$, meaning below the red
solid (dashed) curve for $\Lambda=10$ ($\sqrt{10}$).  The red curve
sweeps upward and to the right as the quark mass $M$, given in
(\protect{\ref{QuarkMass}}), is increased
and encompasses the whole plane as $M\rightarrow\infty$.
The energy loss $dE/dt$ is as for linear drag below the
dotted line and as for radiation in vacuum above the dotted line.
Noting in particular the smallness of the quark masses chosen
in the plot
and the log-scale used on the vertical axis, we see that there is
a large region of parameter space in which the
classical calculation is valid and the energy loss is
as  for acceleration-induced radiation in vacuum.
}
\label{fig5}}

In Fig.~\ref{fig5} we plot
$\Pi\textswab{w}=\Lambda^2$ as a curve in the $(\textswab{w},\gamma)$-plane
for  $\Lambda=\sqrt{10}$ and for $\Lambda=10$.
Our classical calculation is valid below this curve.
As $\Lambda$ increases, the curve sweeps  upward and
to the right.  In Fig.~\ref{fig5} we
also plot the criterion (\ref{eq51}), namely $\textswab{w}=\gamma^{-3/2}$.
Below this curve, we have found that the energy loss $dE/dt$ is
given by the linear drag result (\ref{eq47}); above this curve,
we have found that $dE/dt$ is given by the vacuum radiation result
(\ref{eq48}).
We see that there is a large region of the
$(\textswab{w},\gamma)$-plane in which our calculation is valid
and the vacuum radiation result is obtained.  For example, for any
fixed $v$ we can pick a $\Lambda$ such that upon increasing $\textswab{w}$
the acceleration dominated regime (\ref{eq51}), where $\Pi$ is
given by (\ref{eq48}), is reached long before
the calculation breaks down at $u_c=\Lambda$.
However, if we
first take the $\omega\rightarrow 0$ limit at fixed $v$, and
then study the
result as a function of increasing $v$, we recover the previously
known result for when the linear drag calculation breaks down.
So, at any fixed $\Lambda$ there is always a (perhaps very small)
value of $\textswab{w}$ below which no acceleration-dominated regime
exists with $u_c<\Lambda$. The shape of the red curves in Fig.~\ref{fig5} can
easily be understood:  at small $v$, we see from (\ref{eq48}) that
$\Pi\textswab{w}={\rm constant}$ corresponds to $\textswab{w}\propto 1/v$;
at larger $v$ but where (\ref{eq48}) still applies, we find $\textswab{w}\propto \gamma^{-2}$;
at some still larger $v$, therefore, the red curve must cross the
$\textswab{w}= \gamma^{-3/2}$ curve; beyond this crossover region,
once (\ref{eq47}) applies,
$\Pi\textswab{w}={\rm constant}$ corresponds to a constant $\gamma$, independent
of $\textswab{w}$, and the red curve becomes vertical.

%%%%%%%%%%%%%%%%%%%%%%%%%%%%%%%%%%%%%%%
\section{Speculations About Linear Motion}
\label{sec4}

We have given our conclusions in Section 3.2, and described
the regime in which our calculation is valid in Section 3.3.  In this Section, we
speculate about possible implications of our
results for quarks moving in a straight line.

Consider a quark moving in a straight line at a constant speed $v$.
This means that there is an external force acting on it, and the
external force is doing work at a rate $dE/dt|_{\rm linear\ drag}$
given by (\ref{lineardrag}).  Now, suppose we turn the external
force off.  The quark will decelerate due to the drag force.  Let us calculate
$dE/dt|_{\rm vacuum\ radiation}$ as if the quark were decelerating with this
deceleration in vacuum. For $\vec a$ in the same
direction as $\vec v$, Mikhailov's result (\ref{mikhailovresult}) becomes
\begin{equation}
\frac{dE}{dt}\Biggl|_{\rm vacuum\ radiation}
= \frac{\sqrt{\lambda}}{2\pi}\, \frac{a^2}{(1-v^2)^3}
=  \frac{\sqrt{\lambda}}{2\pi}\,\frac{1}{M^2}\left(\frac{dp}{dt}\right)^2\ ,
\label{linearvacuumradiation}
\end{equation}
where we have used $p=M\gamma v$.\footnote{The
corrections to Mikhailov's result at finite $M$ due to the fact that
the quarks have a nonzero Compton wavelength have been
explored in Ref.~\cite{Chernicoff:2008sa}. We shall always assume that $\Lambda$ is
large enough that these corrections can be neglected.}
At least initially, $dp/dt$ will be
that due to the drag force, namely~\cite{Herzog:2006gh,Gubser:2006bz}
\begin{equation}
\frac{dp}{dt} = - \frac{\pi}{2} \sqrt{\lambda} T^2 \frac{p}{M}\ .
\label{dragdeceleration}
\end{equation}
We now see that the condition that $dE/dt$ due to the vacuum radiation
(\ref{linearvacuumradiation})
caused by the drag-induced deceleration (\ref{dragdeceleration})
be less than $dE/dt$ due
to the drag itself, (\ref{lineardrag}), simplifies considerably and
becomes just
\begin{equation}
\sqrt{\gamma}\ll \sqrt{\gamma_0}\equiv
\frac{ 2 M}{\sqrt{\lambda} T} = \Lambda\ ,
\label{speedlimit}
\end{equation}
where we have used (\ref{QuarkMass}).
Remarkably, this is the same $\gamma$ at which the classical worldsheet
calculation breaks down, as we discussed in Section 3.3.
So, for the case of linear motion, just at the velocity where
the quark is about to enter an acceleration-dominated
regime (by which we mean a regime in which, if the force moving
the quark at constant velocity is turned off, the energy loss
$dE/dt|_{\rm vacuum\ radiation}$ caused by the drag-induced deceleration
would dominate over the energy loss due to
the drag force itself) the calculation breaks down.
In our calculation of a rotating quark, in contrast, we have found
that the acceleration-dominated
regime sets in long before the calculation breaks down. And,
we have furthermore found that in this acceleration-dominated
regime $dE/dt$ turns
out to be just as if the accelerating quark were radiating in vacuum,
even though it is accelerating in a strongly coupled plasma.

Returning to the linear motion that we are focussing on in this Section,
once the external force is turned off, the motion of a quark
that initially has a large momentum in a direction that we
shall term longitudinal can be described by the
Langevin equations~\cite{Moore:2004tg}
\begin{eqnarray} \label{langevin1}
  \frac{d p_L}{ dt} & = & \xi_L (t) - \mu (p_L) p_L, \\
 \frac{d p_i }{dt} & = & \xi_i (t)
 \label{langevin2}
 \end{eqnarray}
with
 \begin{eqnarray}
 \label{re1}
 \vev{\xi_L (t) \xi_L (t')} & = &  \kappa_L (p_L) \delta (t-t')  \\
 \vev{\xi_i (t) \xi_j (t')} & =&   \kappa_T (p_L) \delta_{ij} \delta (t-t')\ ,
  \label{re2}
 \end{eqnarray}
where $p_L$ and $p_i$ are the longitudinal and transverse momentum
of the particle, respectively.
Henceforth, we shall  denote $p_L$ by
$p$.
$\xi_L$ and $\xi_T$ are random fluctuating forces in the
longitudinal and transverse directions. $\vev{\ldots}$ denotes the
medium average.
In general, $\mu$, $\kappa_T$ and
$\kappa_L$ will depend on $p$, but in ${\cal N}=4$ SYM theory
in the regime (\ref{speedlimit}) they are given
by~\cite{Herzog:2006gh,Gubser:2006bz,CasalderreySolana:2006rq,Gubser:2006nz,CasalderreySolana:2007qw}
\begin{equation}
 \label{ne3}
 \mu = \frac{\pi}{2}\sqrt{\lam} \frac{T^2}{M}, \qquad
 \kappa_T = \pi \sqrt{\lam} T^3 \sqrt{\ga}, \qquad
 \kappa_L = \pi \sqrt{\lam} T^3 \ga^{5/2} \ .
\end{equation}
The calculations that lead to (\ref{ne3}) are built upon the
validity of the trailing string solution (\ref{linearstring})
as the description in the gravity dual of the quark moving
at constant velocity, and so are valid only in the regime
(\ref{speedlimit}). However, when  the
velocity reaches the point at which $\gamma\sim\gamma_0$
and (\ref{ne3}) therefore breaks down, there is no sign
that the Langevin description (\ref{langevin1}-\ref{re2}) itself
is breaking down.  To see this, note that the solution of
(\ref{langevin1}) is given by
 \be
 p (t) = \int^t_0 dt' e^{\mu (t'-t)} \xi_L (t') + p(0)e^{-\mu t}\ ,
 \ee
where we have assumed that $\mu$ is $p$-independent and hence
$t$-independent.  From this solution we find that
 \be
 \vev{(\delta p)^2(t)} = \frac{\kappa_L }{2 \mu} \le(1 - e^{-2 \mu t}
 \ri)
 \ee
which means that as long as $t$ is not $\ll 1/\mu$ we have
\be \label{fluc}
 \vev{(\delta p)^2(t)} = \frac{\kappa_L }{ 2 \mu}\ .
\ee
The Langevin description breaks down if the fluctuations (\ref{fluc})
become large compared to $p=M\gamma v$.
We can now see that while (\ref{ne3}) is valid,
\begin{equation}
\frac{ \vev{(\delta p)^2(t)} }{ M^2 \gamma^2 v^2}
= \frac{\sqrt{\gamma}}{v^2} \frac{T}{M} \ .
\label{LangevinBreakdown}
\end{equation}
According to (\ref{speedlimit}), (\ref{ne3}) ceases to be valid
at the velocity
where $\sqrt{\gamma}\sim\sqrt{\gamma_0}$. But, at this
velocity the right-hand side of (\ref{LangevinBreakdown}) is
$\sim 1/\sqrt{\lambda}$, meaning that it is parametrically small.
So, at the velocity at which the trailing string description
and hence the calculation of (\ref{ne3}) breaks down, the
Langevin description remains sound.

So far, we have described facts, not speculations.

We now speculate that a quark moving in a straight
line with  $\gamma>\gamma_0$ is in a regime in which
the Langevin description continues to be valid but the energy loss $dE/dt$
is dominated by the radiation induced by the acceleration
due to the fluctuating forces $\xi_L$ and $\xi_i$ in the Langevin
description.  And, inspired by our results for the rotating quark,
we further speculate that this $dE/dt$ is as if the accelerating quark
were radiating in vacuum, namely
\be
\frac{d E}{dt}\Biggl|_{\rm vacuum\ radiation}
= \frac{\sqrt{\lam}}{2 \pi M^2}
 \le(F_{\parallel}^2 + \ga^2 \vec F_{\perp}^2 \ri)
\ ,
\label{rea}
\ee
where $F_\parallel^2$ receives contributions from drag and
from the fluctuating forces while $F_\perp^2$ is due to
the fluctuating forces alone.
Near $\gamma\sim\gamma_0$, where (\ref{ne3}) may still
be a qualitative guide, perhaps the ratio
of the contribution of the fluctuating forces to $F_\parallel^2$
and $F_\perp^2$ is of order $\gamma^2$. If
so, then the two terms on the right-hand side
of (\ref{rea}) make comparable contributions.
It would be interesting to
determine whether
the $\gamma^2  F_\perp^2$ term dominates when $\gamma\gg \gamma_0$, as
is the case in a weakly coupled plasma.

We can now clearly see the significant advantage offered by
the analysis of the rotating quark. In that situation, the
onset of the acceleration-dominated regime occurs when the
analogue of (\ref{speedlimit}) is satisfied, meaning that we
can do a reliable calculation.  Upon so doing, we reached the
conclusions described in Section 3, and in particular discovered
that the energy loss in the acceleration-dominated regime is
as if the quark were in vacuum.

\medskip
\section*{Acknowledgments}
\medskip

We acknowledge helpful conversations with Christiana Athanasiou,
Tom Faulkner, Misha Stephanov and Laurence Yaffe.
KBF is supported in part by Shahrood
University of Technology research grant No. 24015 and by CERN.
HL is supported in part by the A.~P.~Sloan Foundation and the U.S.
Department of Energy (DOE) OJI program. 
This research was
supported in part by the DOE Offices of Nuclear and High Energy
Physics under grants \#DE-FG02-94ER40818 and \#DE-FG02-05ER41360.

%%%%%%%%%%%%%%%%%%%%%%%%%%%%%%%%%%%%%%%%%%%%%%%%%%%%%%%%%%%%%%%%%%%%%%%%%%%%%%%%

\end{document}